\documentclass[aps,prc,twocolumn,superscriptaddress,showpacs,showkeys]{revtex4}

\usepackage{graphicx,amsmath,amssymb,bm}
\newcommand{\eq}[1]{Eq.~(\ref{#1})}
\newcommand{\be}{\begin{equation}}
\newcommand{\ee}{\end{equation}}
\newcommand{\bea}{\begin{eqnarray}}
\newcommand{\eea}{\end{eqnarray}}
\newcommand{\vnn}{V_{\rm NN}}
\newcommand{\la}{\langle}
\newcommand{\ra}{\rangle}
\newcommand{\tr}{{\rm Tr}}
\newcommand{\fm}{\, \text{fm}}
\newcommand{\fmi}{\, \text{fm}^{-1}}
\newcommand{\mev}{\, \text{MeV}}

\newcommand{\dxsect}{\frac{d\sigma}{d\Omega}}
\newcommand{\mj}{m_j}
\newcommand{\mjp}{m_j'}
\newcommand{\mn}{m_n}
\newcommand{\mnp}{m_n'}
\newcommand{\q}{{\bm q}}
\newcommand{\qp}{{\bm q'}}
\newcommand{\qhat}{\widehat{\bm q}}
\newcommand{\qphat}{\widehat{\bm q'}}
\newcommand{\p}{{\bm p}}
\newcommand{\pp}{{\bm p'}}
\newcommand{\phat}{\widehat{\bm p}}
\newcommand{\pphat}{\widehat{\bm p'}}
\newcommand{\mdag}{M^\dagger}
\newcommand{\sm}{\bm \sigma}
\newcommand{\phist}{\widetilde{\phi}_d^0}
\newcommand{\phidt}{\widetilde{\phi}_d^2}
\newcommand{\phiop}{\widehat{\phi}_d}
\newcommand{\St}{\Sigma}
\newcommand{\Stp}{\Sigma'}
\newcommand{\mSt}{m_\Sigma}
\newcommand{\mStp}{m_\Sigma'}
\newcommand{\lam}{\lambda}
\newcommand{\lamp}{\lambda'}
\newcommand{\mlam}{m_\lambda}
\newcommand{\mlamp}{m_\lambda'}
\newcommand{\pa}{{\bm p}_1}
\newcommand{\pb}{{\bm p}_2}
\newcommand{\pc}{{\bm p}_3}
\newcommand{\pap}{{\bm p'_1}}
\newcommand{\pbp}{{\bm p'_2}}
\newcommand{\kk}{{\bm k}}
\newcommand{\kkp}{{\bm k'}}
\newcommand{\pt}{{\bm P}}
\newcommand{\sia}{{\bm \sigma}_1}
\newcommand{\sib}{{\bm \sigma}_2}
\newcommand{\sic}{{\bm \sigma}_3}
\newcommand{\vo}{v_{\openone}}
\newcommand{\vsp}{v_{\rm spin}}
\newcommand{\bigvo}{V_{\openone}}
\newcommand{\bigvsp}{V_{\rm spin}}

\newcommand{\ord}{{\mathcal O}}
\newcommand{\si}{{\bm \sigma}_3}
\newcommand{\sd}{{\bm S}}
\newcommand{\er}{E_{\rm rel}}

\begin{document}

\preprint{NT@UW-07-04}

\title{Resonant relativistic corrections and the $A_y$ problem}

\author{G.A.~Miller}
\email[E-mail:~]{miller@phys.washington.edu}
\affiliation{Department of Physics, University of Washington,
Seattle, WA 98195}
\author{A.~Schwenk}
\email[E-mail:~]{schwenk@triumf.ca}
\affiliation{Department of Physics, University of Washington,
Seattle, WA 98195}
\affiliation{TRIUMF, 4004 Wesbrook Mall, Vancouver, BC, Canada, V6T 2A3}

%\date{\today}

\begin{abstract}
We study relativistic corrections to nuclear interactions caused by
boosting the two-nucleon interaction to a frame 
in which their total momentum does not vanish. These corrections induce a 
change in the computed value of the neutron-deuteron analyzing power $A_y$
that is estimated using the plane-wave impulse 
approximation. This allows a transparent analytical calculation that
demonstrates the significance of relativistic corrections.
Faddeev calculations are however needed to conclude on the $A_y$ puzzle.
\end{abstract}


\maketitle

\section{Introduction}

One of the major unsolved problems in nuclear physics is the so-called 
$A_y$ puzzle in nucleon-deuteron (N$d$) scattering. 
The nucleon analyzing power 
$A_y$ is the difference in differential cross sections for scattering 
of polarized nucleons~\cite{Seyler}:
\be
A_y = 
\frac{\dxsect\bigr|_{\uparrow} 
- \dxsect\bigr|_{\downarrow}}{\dxsect\bigr|_{\uparrow} 
+ \dxsect\bigr|_{\downarrow}} \,,
\label{Aydiff}
\ee
where $\uparrow$ denotes the polarization normal to the reaction plane
(spanned by the center-of-mass momentum of the incident and scattered
nucleon). All modern nucleon-nucleon (NN) interactions lead to 
practically the same results: They under predict $A_y$ by $30\%$ 
for laboratory energies $E_{\rm N} \lesssim 30 \mev$ (for a review 
see~\cite{Gloeckle}), whereas the predicted $A_y$ is in very good 
agreement for higher energies. The contributions of the existing 
three-nucleon (3N) interactions to $A_y$ are small at low 
energies~\cite{Gloeckle,Kievsky,chiral3NF}. A similar discrepancy 
is found for the deuteron vector analyzing power 
$i T_{11}$~\cite{Gloeckle}.

The NN contribution to $A_y$ is directly related to the $^3$P$_j$ phase 
shifts~\cite{PwavesWitala}, but it is very unlikely that uncertainties 
in these phases can resolve the puzzle~\cite{PwavesFriar}. For few $\mev$ 
energies, $A_y$ is maximal around a center-of-mass scattering angle 
$\theta \approx 100^\circ$. This is the
location of the minimum of the differential 
cross section so that small effects are amplified in $A_y$. As 
a result, a number of small contributions to $A_y$ have been investigated. 
For instance, magnetic moment interactions lead to a very small 
contribution to $A_y$ near the maximum in $pd$ scattering, but are only 
sizable at forward angles for $nd$~\cite{Stoks,KievskyMM}. Moreover, ad hoc
solutions have been proposed that range from introducing a phenomenological
3N spin-orbit force~\cite{KievskyLS} to including the effects of exchanging 
one pion in the presence of a two-nucleon correlation~\cite{Canton}.

Recently, Fisher {\it et al.} have shown that the $A_y$ problem
increases from a $30\%$ discrepancy in N$d$ to 
a $100\%$ puzzle in $p \, ^3$He~\cite{p3He}. Therefore, one
can expect that the problem becomes even more pronounced for
understanding heavier systems. In addition, the $A_y$ discrepancy
increases with the inclusion of the Coulomb interaction in the
$pd$ system~\cite{Coulomb1,Coulomb2,Coulomb3}.

In the three-body (or higher-body) system, not all pairs of
particles are simultaneously in the {\it two-body} center-of-mass 
(cm) system, and therefore relativistic 
corrections~\cite{rel1,rel2,rel3} have to be taken into account:
\be
\delta V \sim \frac{Q^2}{m^2} \: \vnn \,,
\ee
where $Q^2$ includes at least one power of
the two-body cm momentum ${\bm P} = {\bm p_1}+{\bm p_2}$, 
$m$ is the nucleon mass and $\vnn$ denotes the NN interaction in the 
cm frame (for $P=0$). The modern understanding is to consider
these corrections as 3N interactions, but using the formalism
developed in~\cite{rel1,rel2,rel3} it is straightforward to include
these effects to order $(Q/m)^2$ without any new parameters. The
naive expectation is that relativistic corrections are small
at low energies. This was confirmed for selected $nd$ observables
and for energies $E_n \geqslant 28 \mev$~\cite{rel_nd} (where
there is no $A_y$ problem).

In this work, we show that, in contrast to the naive expectation,
relativistic boost effects may be important at low energies. 
This is due to spin-violating relativistic corrections, which 
couple relative NN S-waves with the $^3$P$_j$ waves (combined 
with a change of the two-body cm angular momentum). We find that
the interference with the large S-wave scattering lengths can
lead to resonant enhancements of $A_y$ at low energies. This
effect would explain why predictions for $A_y$ at $E_n \gtrsim
30 \mev$ agree well with experiment. We present a transparent 
analytical calculation, based on using the plane-wave impulse
approximation, that explores the effect of relativistic corrections on 
spin observables. The effects are small but significant and should be
combined with a complete solution of the Faddeev equations.

This paper is organized as follows. We begin in Section~\ref{Notation} with
a brief discussion of the relevant notation and scattering formalism.
In Section~\ref{RelCorr},
we classify all relativistic corrections to order $(Q/m)^2$
and their impact on the differential cross section and $A_y$. 
We calculate analytically their effect on $A_y$ neglecting
distortions. The central findings of this paper are given in 
Eq.~(\ref{dUsv}) and in Fig.~\ref{Ay_ndE3}. Our results and 
Eq.~(\ref{dUsv}) are general and in a form that should be 
implemented in future Faddeev calculations. The reader familiar with
the standard notation and 3N scattering can skip Section~\ref{Notation}.
%Our results are given in a form implementable in
%Faddeev calculations.
In Section~\ref{Results}, the contribution to 
$A_y$ is estimated using benchmarked $nd$~\cite{Kievsky,KievskyPS} 
phase shifts and pionless effective field theory (EFT) 
contact interactions for the relativistic corrections $\delta V$. 
We conclude in Section~\ref{Concl}
that relativistic boost effects may be important for a precise 
understanding of three-body spin observables.

\section{Notation and Scattering Formalism}
\label{Notation}

We follow the notation and conventions of Gl\"ockle {\it et 
al.}~\cite{Gloeckle} and define the $nd$ scattering amplitude $M$
by
\begin{multline}
M_{\mjp,\mnp;\mj,\mn}(\qp,\q) \\[1mm]
= - \frac{2m}{3} \, (2\pi)^2 \,
\la \phi_d , \mjp ; \qp , \mnp | U | \phi_d , \mj ;
\q , \mn \ra \,,
\label{MU}
\end{multline}
where $\mj, \mn$ are the deuteron total angular momentum
and nucleon spin magnetic quantum numbers respectively, $\q, \qp$ 
are initial and final relative momenta of the nucleon in the 
$nd$ cm system, and $U$ denotes the transition amplitude.
The relative momenta are on-shell 
related to the neutron laboratory energy $E_n$ by $q = 
|\q|=|\qp|= \sqrt{\frac{8}{9} \, m \, E_n}$, and the cm 
scattering angle is $\cos\theta \equiv \qhat \cdot \qphat$. Finally, 
the plane-wave states are normalized as $\la \pp | p \, l \,
m \ra \equiv i^{-l} \, Y_{lm}\bigl(\pphat\bigr) \, \delta(p'-p)/(p'p)$
with spherical harmonics $Y_{lm}(\phat)$, and thus
$\la \p | \pp \ra = \delta^{(3)}(\p-\pp)$ and $\la \p | {\bm r} \ra
= e^{-i \p \cdot {\bm r}}/(2 \pi)^{3/2}$.

In terms of the scattering amplitude, the spin-averaged differential 
cross section $d\sigma/d\Omega$ is given by
\begin{align}
\frac{d\sigma}{d\Omega} &= \frac{1}{(2j+1)(2 s_{\rm N} +1)} \: \tr \bigl(
M \mdag \bigr) \nonumber \\[1mm]
&= \frac{1}{6} \sum_{\mjp,\mnp,\mj,\mn} 
\bigl| M_{\mjp,\mnp;\mj,\mn}(\qp,\q) \bigr|^2 \,,
\end{align}
where $j=1$ and $s_{\rm N}=1/2$ are the spin of the deuteron and nucleon
respectively. The nucleon analyzing power is defined by
\be
A_i = \frac{\tr \bigl( M \sm^i \mdag \bigr)}{\tr \bigl( M \mdag \bigr)} \,,
\ee
with Pauli matrices $\sm^i$ and standard conventions for the
coordinate system: $\widehat{\bf z} = \qhat$, $\widehat{\bf y} =
\qhat \times \qphat/|\qhat \times \qphat|$ and $\widehat{\bf x} =
\widehat{\bf y} \times \widehat{\bf z}$. This directly leads to
Eq.~(\ref{Aydiff}), if one chooses $\widehat{\bf y}$ as the spin 
quantization axis. Finally, due to parity conservation, $A_x = 
A_z = 0$~\cite{Seyler}. Using the Fourier transform of 
an operator representation for the
deuteron wave function~\cite{BrownJackson}
\begin{align}
\phiop(\p) \equiv \phist(p) + \phidt(p) \, \frac{S_{12}(\phat)}{\sqrt{8}} 
\,, \\[1mm]
\la \p , \mjp | \phi_d , \mj \ra = \bigl\langle \mjp \bigl| \,
\phiop(\p)  \, \bigr| \mj \bigr\rangle \,,
\end{align}
with tensor operator $S_{12}(\phat)$, the scattering amplitude 
can be expressed in a convenient operator form
\begin{widetext}
\be
M_{\mjp,\mnp;\mj,\mn}(\qp,\q)
= - \frac{2m}{3} \, (2\pi)^2 \int d\pp \int d\p \:
\bigl\langle \pp , \mjp ; \qp , \mnp \bigl| \, \phiop(\pp) \, U \, 
\phiop(\p) \, \bigr| \p , \mj ; \q , \mn \bigr\rangle \,.
\label{MUop}
\ee
\end{widetext}
Finally, computing the analyzing power is simplified by 
coupling the deuteron total angular momentum ${\bm j}$ with the nucleon
spin ${\bm s}_{\rm N}$ to a total spin ${\bm \Sigma} = {\bm j} +
{\bm s}_{\rm N}$. In this basis, the spin matrix elements of the
scattering amplitude are given by
\begin{widetext}
\be
M_{\Stp,\mStp;\St,\mSt}(\qp,\q) = \sum_{\mjp,\mj} 
(1 \, \mjp \, 1/2 \, \mStp-\mjp | \Stp \, \mStp )
(1 \, \mj \, 1/2 \, \mSt-\mj | \St \, \mSt )
M_{\mjp,\mStp-\mjp;\mj,\mSt-\mj}(\qp,\q) \,.
\ee
\end{widetext}

We use benchmarked $nd$ partial waves for the scattering 
amplitude without relativistic corrections, so we briefly discuss
the partial wave expansion. The states with
good total spin $\St$ read $| p \, (ls) j ; \q \, (j \, 1/2) \St 
\, \mSt \ra$, where $s=1$ is the spin of the deuteron and the
nucleon motion can also be expanded in angular momenta $| q \,
\lam \, \mlam \ra$. In these states the $nd$ scattering
amplitude is given by
\begin{widetext}
\begin{multline}
M_{\Stp,\mStp;\St,\mSt}(\qp,\q) = - \frac{2m}{3} \, (2\pi)^2 
\sum_{\lamp,\mlamp,\lam,\mlam} i^{\lam-\lamp} \, Y_{\lamp,\mlamp}(\qphat)
\, Y^*_{\lam,\mlam}(\qhat) \\[1mm]
\times \sum_{l,l'} \int p'^2 dp' \, \widetilde{\phi}_d^{l'}(p')
\int p^2 dp \, \widetilde{\phi}_d^l(p) \,
\bigl\langle p' \, (l' 1) 1 ; q' \, \lamp \, \mlamp \, (1 , 1/2) 
\Stp \, \mStp \bigl| \, U \, \bigr| p \, (l 1) 1 ; q \, \lam \, \mlam \, 
(1 , 1/2) \St \, \mSt \bigr\rangle \,.
\label{Mpw1}
\end{multline}
\end{widetext}
Next one couples the nucleon angular momentum with the total spin 
to the total angular momentum ${\bm J} = {\bm \lambda} +
{\bm \Sigma}$, for which $U$ is diagonal in $J$ and independent
of $m_J$, thus $\mlamp + \mStp = \mlam + \mSt$. With $\qhat =
\widehat{\bf z}$, we have $Y^*_{\lam,\mlam}(\qhat)=\delta_{\mlam,0} \,
\sqrt{\frac{2\lam+1}{4\pi}}$, and consequently $m_J = m_\Sigma$
and $\mlamp = \mSt - \mStp$. The second line in Eq.~(\ref{Mpw1}) 
in the coupled $(\lam \Stp) J \, m_J$ basis is independent of $m_J$
and can be decomposed as $(\delta_{\lamp,\lam} \delta_{\Stp,\St}
- S_{\lamp,\Stp;\lam,\St}^J)/(4 \pi i m q/3)$. With this at hand,
the partial wave decomposition reads
\begin{widetext}
\begin{align}
M_{\Stp,\mStp;\St,\mSt}(\qp,\q) &= \frac{i \sqrt{\pi}}{q}
\sum_{\lamp,\lam,J} i^{\lam-\lamp} \sqrt{2\lam+1} \:
Y_{\lamp,\mSt-\mStp}(\qphat) \nonumber \\[1mm]
&\times ( \lamp \, \mSt-\mStp \, \Stp \, \mStp | J \, \mSt )
\, ( \lam \, 0 \, \St \, \mSt | J \, \mSt ) \,
\bigl( \delta_{\lamp,\lam} \delta_{\Stp,\St}
- S_{\lamp,\Stp;\lam,\St}^J \bigr) \,,
\label{Mpw2}
\end{align}
\end{widetext}
where $S_{\lamp,\Stp;\lam,\St}^J$ is given in terms of the $nd$
phase shifts and mixing parameters~\cite{Seyler} (see also
Eqs.~(209)--(214) in Ref.~\cite{Gloeckle}).

\section{Relativistic corrections}
\label{RelCorr}

Boost corrections to the two-nucleon interaction depend on the 
total momentum $\pt$ of the pair and are obtained by 
satisfying the commutation relations of the Poincar\'{e} group~\cite{rel1}.
To leading order in $(Q/m)^2$, the relativistic boost corrections are 
given in momentum space by (for the corresponding coordinate space
expression, see Eq.~(1.7) in Ref.~\cite{rel3})
\begin{widetext}
\begin{align}
\delta v_{\sia,\sib}(\kkp,\kk,\pt) &= -\frac{P^2}{4m^2} \,
v_{\sia,\sib}(\kkp,\kk)
+ \frac{i}{8 m^2} \, \bigl[ (\sia - \sib) , v_{\sia,\sib}(\kkp,\kk) \bigr]
\times \pt \cdot \kk \nonumber \\[1mm]
&- \frac{i}{8 m^2} \, (\sia - \sib) \times \pt \cdot (\kk - \kkp) \,
v_{\sia,\sib}(\kkp,\kk)
- \frac{1}{8 m^2} \, \bigl( \pt \cdot (\kk - \kkp) \bigr) \, \pt \cdot
\nabla_{\kk-\kkp} \, v_{\sia,\sib}(\kkp,\kk) \,,
\label{dVdirect}
\end{align}
\end{widetext}
where $v_{\sia,\sib}(\kkp,\kk,\pt)$ is the direct NN interaction in
the cm system, with initial and final relative momenta $\kk = (\pa - 
\pb)/2$ and $\kkp = (\pap - \pbp)/2$, and Eq.~(\ref{dVdirect}) accounts 
only for the direct term of the boost correction.
As explained in Ref.~\cite{rel3} the Poincar\'{e} group commutation 
relations do not have a unique solution. The operator $\delta v$ can 
have an additional term of the form 
\be 
\delta v' = -i \, \bigl[ \chi , H_0 + v \bigr] \,,
\label{extra}
\ee
where $\chi$ is a translationally invariant function and 
$H_0$ is the non-interacting Hamiltonian. One must pay attention 
to this term when studying scattering processes.

Some of the operators in Eq.~(\ref{dVdirect}) can be obtained from purely
classical considerations~\cite{rel3}. The first term arises from treating
the potential as a contribution to the nucleon mass and then
expanding the relativistic energy operator. The final term of  
Eq.~(\ref{dVdirect}) is due to the effects of Lorentz contraction. 
The third term results from Thomas precession in which objects with 
spin precess when they accelerate, since rotations do not commute 
with boosts. The commutator term of Eq.~(\ref{dVdirect})
does not have an analog in classical mechanics.

Our procedure will be to calculate the leading relativistic corrections 
$\delta M$ in the basis $| \p , \mj ; \q , \mn \ra$ of Eq.~(\ref{MUop})
by accounting for the change in $U$, $\delta U$ caused by $\delta v$ 
of Eq.~(\ref{dVdirect}). In this exploratory study we use the 
plane-wave impulse approximation, which treats one of the nucleons
as a spectator. Taking the matrix element of $\delta U$ 
within plane-wave neutron-deuteron states yields $\delta M$. The use of 
the plane-wave impulse 
approximation enables us to analytically study the effect of
relativistic boost corrections and make a first assessment of their
importance. A full Faddeev calculation including distortions will 
eventually be needed to make a complete assessment. 
Our present use of the plane-wave impulse approximation has an additional 
advantage: The matrix element of the term $\delta v'$ of \eq{extra}, 
taken between on-shell elastic scattering states vanishes.

There are three contributions to $\delta U$ arising from the 
three pairs in the $nd$ system:
\be
\delta U = \delta V_{12} + \delta V_{13} + \delta V_{23} \,.
\ee

Since relativistic corrections are of order $(Q/m)^2$, we keep
only the central parts in $v_{\sia,\sib}(\kkp,\kk)$. Non-central
interactions start at $\ord\bigl((Q/m_\pi)^2\bigr)$ in pionless effective
field theory, which is relevant for the energies of interest. 
Therefore, the spin structure is limited to 
\be 
v_{\sia,\sib} = \vo + \vsp \, \sia \cdot \sib \,.
\ee
Furthermore we can neglect the last term in 
Eq.~(\ref{dVdirect}) of $\ord\bigl((Q/m)^2 (Q/m_\pi)^2\bigr)$.

The next step is to include the exchange term. We need to compute
\begin{multline}
\delta V_{\sia,\sib}(\kkp,\kk,\pt) = \\[1mm]
\delta v_{\sia,\sib}(\kkp,\kk,\pt) - P_{\bm \sigma} P_{\bm \tau} \:
\delta v_{\sia,\sib}(-\kkp,\kk,\pt) \,,
\end{multline}
where the spin (isospin) exchange operator is  $P_{\bm \sigma}$ 
($P_{\bm \tau}$) and $V_{\sia,\sib}(\kkp,\kk) = v_{\sia,\sib}(\kkp,\kk) 
- P_{\bm \sigma} P_{\bm \tau} \: v_{\sia,\sib}(-\kkp,\kk)$ denotes
the antisymmetrized interaction. Writing the commutator term of 
Eq.~(\ref{dVdirect}) explicitly and using the property that 
$\{ (\sia - \sib) , P_{\bm \sigma} \} = 0$ leads directly to the 
result:
\begin{widetext}
\be
\delta V_{\sia,\sib}(\kkp,\kk,\pt) = -\frac{P^2}{4m^2} \,
V_{\sia,\sib}(\kkp,\kk)
- \frac{i}{8 m^2} \, V_{\sia,\sib}(\kkp,\kk) \, (\sia - \sib) 
\times \pt \cdot \kk 
+ \frac{i}{8 m^2} \, (\sia - \sib) \times \pt \cdot \kkp \,
V_{\sia,\sib}(\kkp,\kk) \,.
\label{dVdirplusex}
\ee
\end{widetext}
In antisymmetrized states, relativistic corrections thus have the
form of $V$~(boost corrections in)$-$(boost corrections out)~$V$.

It is evident that the first ($P^2$) term in Eq.~(\ref{dVdirplusex}) 
will lead to a relativistic correction to the $nd$ scattering amplitude 
that is a scalar in spin and of general structure:
\be
\delta M_{P^2} \sim \openone \: \frac{Q^2}{m^2} \quad {\rm and} \quad
\sd \cdot \si \: \frac{Q^2}{m^2} \,,
\ee
where $\sd = ( \sia + \sib )/2$ is the deuteron spin and
the nucleon spin operator is given by ${\bm s}_{\rm N} = \sic/2$.
In the following we will show that the spin-violating (sv)
relativistic corrections (the last two terms in 
Eq.~(\ref{dVdirplusex})) lead to terms of the form
\be
\delta M_{\rm sv} \sim \sd^y \: \frac{Q^2}{m^2} \quad
{\rm and} \quad \si^y \: \frac{Q^2}{m^2} \,.
\ee
The leading contributions to the differential cross section and
to $A_y$ are from the interference of $\delta M$ of $\ord\bigl(
(Q/m)^2 \bigr)$ with the leading $nd$ scattering amplitude at 
low energies. Similar to the above considerations for the two-nucleon
interaction, the leading operators in $M$ are given by the central
part:
\be
M_{\sd,\si} = M_{\openone} + M_{\rm spin} \, \sd \cdot \si + 
\ord\bigl((Q/m_\pi)^2\bigr) \,.
\ee
We can now evaluate the relativistic corrections to the nucleon analyzing 
power $\delta A_y$ and to the differential cross section
$\delta (d\sigma/d\Omega)$:
\begin{align}
\delta A_y &= \frac{\tr \bigl( \delta M \, \sm^y \mdag + M \sm^y \,
\delta \mdag \bigr)}{\tr \bigl( M \mdag \bigr)} \nonumber \\[1mm]
&- A_y \: \frac{\tr \bigl( \delta M \, \mdag + M \, \delta \mdag \bigr)}
{\tr \bigl( M \mdag \bigr)} \,,
\label{dAygen} \\[2mm]
\delta \frac{d\sigma}{d\Omega} &= \frac{1}{6} \:
\tr \bigl( \delta M \, \mdag + M \, \delta \mdag \bigr) \,.
\label{dcrossgen}
\end{align}
Since $A_y$ is small, we can neglect the second term in 
Eq.~(\ref{dAygen}). For the leading contributions, it then follows
that only $\delta M_{\rm sv}$ contributes to $\delta A_y$,
\be
\delta A_y = \frac{\tr \bigl( \delta M_{\rm sv} \, \sm^y \mdag + M \sm^y \,
\delta \mdag_{\rm sv} \bigr)}{\tr \bigl( M \mdag \bigr)} +
\ord\biggl( \frac{Q^4}{m^2 m_\pi^2} \biggr) \,,
\ee
and only $\delta M_{P^2}$ contributes to $\delta (d\sigma/d\Omega)$.
A straightforward calculation of the spin-violating relativistic 
corrections arising from $\bigvo(\kkp,\kk) + \bigvsp(\kkp,\kk) \, 
\sia \cdot \sib \equiv (1-P_{\bm \sigma} P_{\bm \tau} P_{\bm k})
(\vo + \vsp \, \sia \cdot \sib)$ yields
\begin{widetext}
\be
\delta V^{\rm sv}_{\sia,\sib}(\kkp,\kk,\pt) = 
- \frac{i}{8 m^2} \, (\sia - \sib) \times \pt \cdot (\kk-\kkp)
\, \bigl( \bigvo(\kkp,\kk) -
\bigvsp(\kkp,\kk) \bigr)
+ \frac{1}{4 m^2} \, (\sia \times \sib) \times \pt \cdot 
(\kk + \kkp) \, \bigvsp(\kkp,\kk) \,.
\label{dVsv}
\ee
\end{widetext}
The resulting spin-violating interactions connect the two-body 
$^3$P$_j$ waves (which are crucial for $A_y$) with the two-body
S-waves. The S-waves are resonant at low energies with large
scattering lengths, $a_0 \equiv a_{^1{\rm S}_0} = - 23.768 \pm 0.006 \fm$ 
and $a_1 \equiv a_{^3{\rm S}_1} = 5.420 \pm 0.001 \fm$~\cite{as}, 
and therefore
the interference with the $^3$P$_j$ waves can lead to a resonant
enhancement of these relativistic corrections at low-energies.
For higher energies, the S-wave phase shifts decrease, so
the effect of the spin-violating interactions decreases.

Including isospin and restricting two-nucleon interactions to 
S-waves, the central part of the 
antisymmetrized two-body interaction can be written as
\begin{multline}
V_{i,3} = \frac{1}{8} \bigl[ V_0 \, (1-{\bm \sigma}_i \cdot {\bm \sigma}_3) 
(3+ {\bm \tau}_i \cdot {\bm \tau}_3) \\[1mm]
+ V_1 \, (3+{\bm \sigma}_i \cdot {\bm \sigma}_3) (1- {\bm \tau}_i
\cdot {\bm \tau}_3)\bigr] \,,
\end{multline}
where $i=1,2$ and ${\bm \tau}_{i,3}$ denote Pauli matrices that operate in
isospin space and $V_{0,1}$ are projections on $s=0,1$ states.
The operator ${\bm \tau}_i \cdot {\bm \tau}_3$ vanishes in $nd$
states, and thus we have
\be
\bigvo = \frac{3}{8} \, (V_0 + V_1) \quad {\rm and} \quad
\bigvsp = \frac{1}{8} \, (V_1 -3 V_0) \,.
\ee

We use leading-order $(Q/m_\pi)^0$ pionless EFT contact 
interactions~\cite{ksw}, where the operators $V_0$ and 
$V_1$ are momentum independent:
\be
V_i = \frac{C_i}{2 \pi^2 m} \quad {\rm with} \quad
C_i = \frac{1}{\frac{1}{a_i} - \mu} \,,
\label{Bornc}
\ee
for $i=0,1$. Here, $\mu$ is the renormalization
scale in dimensional regularization with power-divergence
subtraction scheme~\cite{ksw}. A similar expression is
obtained for a momentum-cutoff regularization.
The operator of \eq{dVsv} therefore has the form of a 
spin-violating operator ($\sia - \sib$) dotted into a 
momentum vector that induces transitions between spin triplet 
(singlet)/relative S-wave and spin singlet (triplet)/relative 
P-wave states. The momentum vector $\kk$ ($\kkp$) in \eq{dVsv}
explicitly projects on incoming (outgoing) P-wave states.
The change in the orbital angular momentum is compensated
with a corresponding change of the two-body cm angular momentum,
so that the total angular momentum is preserved.

At low energies we can take the P-wave states to be plane waves,
but we include iterated S-wave interactions $V_i$ in the initial
state (for $\kkp$) and final state (for $\kk$). This leads to
replacing $C_i$ by
\be
C_i \to \frac{C_i}{1 + C_i \, (\mu +i \sqrt{m \, \er} )} 
= \frac{1}{\frac{1}{a_i} + i \sqrt{m \, \er}} \,,
\label{newc}
\ee
where $\er$ is the relative energy (in the two-body cm of system).
As a result of these initial and final state interactions, we
find that the operators $V_i$ that enter in \eq{dVsv} are
independent of the renormalization scale $\mu$.

\begin{figure}[t]
\begin{center}
\includegraphics[scale=0.7,clip=]{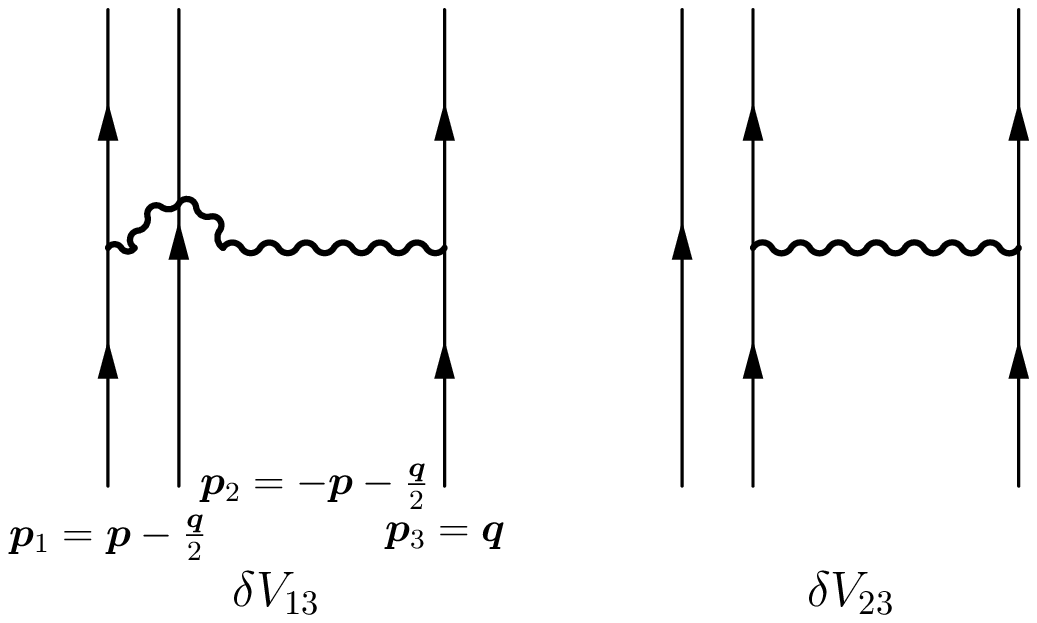}
\caption{Contributions of the spin-violating interactions to the
relativistic corrections $\delta U_{\rm sv}$
and our conventions for the Jacobi momenta. In Born approximation,
the low-energy coefficients are given by the $C_i$ of \eq{Bornc}, and
in the plane-wave impulse approximation we use the $C_i$ of \eq{newc}.}
\label{diags}
\end{center}
\end{figure}

Next we calculate the contributions of the spin-violating relativistic
corrections to $\delta A_y$ for the $nd$ system (where the
Coulomb interaction does not operate). We neglect distortions
that would involve the nucleon treated as a spectator. 
We work in the
three-body cm system $\pa + \pb + \pc = 0$, where nucleons $1$, $2$ 
constitute the deuteron, and nucleon $3$ is the free neutron.
Since the matrix element of $\delta V^{\rm sv}_{12}$ vanishes 
when evaluated in the deuteron eigenstate, we need only to evaluate
two contributions to the spin-violating collision operator 
$\delta U_{\rm sv} = \delta V^{\rm sv}_{13} + \delta V^{\rm sv}_{23}$,
shown diagrammatically in Fig.~\ref{diags}. We employ
Jacobi momenta and the incoming nucleon momenta are expressed as
$\pa = \p - \q/2$, $\pb = -\p -\q/2$ and $\pc = \q$ (with 
primed momentum labels for the outgoing nucleons). The second
contribution $\delta V^{\rm sv}_{23}$ can be obtained from
$\delta V^{\rm sv}_{13}$ by replacing $\sia \to \sib$,
$\p \to - \p$ and $\pp \to - \pp$. Since the deuteron is even
in momentum ($l=0,2$), we can change variables in Eq.~(\ref{MUop})
back to $-\p \to \p$ and $-\pp \to \pp$. Consequently, the
contribution of $\delta V^{\rm sv}_{23}$ to $\delta M_{\rm sv}$
is identical to the contribution of $\delta V^{\rm sv}_{13}$
after replacing $\sia \to \sib$ in the latter.

Inserting this into Eq.~(\ref{dVsv}) (with $1,2 \to 1,3$ and $1,2
\to 2,3$), we obtain for the total leading $(Q/m)^2$ relativistic 
corrections relevant for $\delta A_y$
\be
\delta U^{\rm sv}_{\sd,\si}(\p,\qp,\q) =
\delta^{(3)}\bigl(\pp - (\p + \Delta) \bigr) \,
\delta \widetilde{U}^{\rm sv}_{\sd,\si}(\p,\qp,\q) \,,
\ee
where the delta function accounts for the conservation of the
two-body cm momentum, the momentum transfer is
$\Delta \equiv (\q-\qp)/2$, and we have
\begin{widetext}
\be
\delta \widetilde{U}^{\rm sv}_{\sd,\si}(\p,\qp,\q)
= \biggl[ - \frac{i}{4 m^2} \, (\sd - \si) \times (\p+\frac{\q}{2}) 
\cdot (\qp-\q) \, \frac{3 V_0 + V_1}{4}
- \frac{1}{2 m^2} \, (\sd \times \si) \times (\p+\frac{\q}{2})  \cdot 
(\q + \qp) \, \frac{V_1 -3 V_0}{8} \biggr] \,.
\label{dUsv}
\ee
\end{widetext}
Here the relative momentum arguments
of $V_{0,1}$ are $\kk = \frac{\p}{2} - \frac{3\q}{4}$ and $\kkp = 
\frac{\p}{2}+\frac{\q}{4} - \qp$.
The result of Eq.~(\ref{dUsv}) is  general and useful as input to
Faddeev calculations, in which the term is dressed by the effects of
initial and final state strong interactions.

Using \eq{dUsv} in Eq.~(\ref{MUop}), we obtain our final expression 
for the relevant change in the scattering amplitude:
\begin{widetext}
\be
\delta M^{\rm sv}_{\mjp,\mnp;\mj,\mn}(\qp,\q)
= - \frac{2m}{3} \, (2\pi)^2 \int d\p \:
\bigl\langle \mjp , \mnp \bigl| \, \phiop(\p + \Delta) \,
\delta \widetilde{U}^{\rm sv}_{\sd,\si}(\p,\qp,\q) \,
\phiop(\p) \, \bigr| \mj , \mn \bigr\rangle \,.
\label{MUsv}
\ee
\end{widetext}
Next, we estimate the impact of these spin-violating boost
corrections on $A_y$ based on pionless EFT contact interactions 
for $V_{0,1}$ and using benchmarked $nd$ phase shifts from 
Kievsky {\it et al.}~\cite{Kievsky,KievskyPS} for $M$. This has the
advantage that $\delta A_y$ can be evaluated analytically and the
physics is transparent.

\section{Results}
\label{Results}

We can transform variables in Eq.~(\ref{MUsv})
from $\p \to \p - \Delta/2$. For momentum-independent interactions
$V_{0,1}$, terms linear in $\p$ in $ \widetilde{U}^{\rm sv}$ integrate 
to zero after this variable transformation. Therefore, we can replace
$\p + \q/2$ by $- \Delta/2 + \q/2 = (\q + \qp)/4$ in Eq.~(\ref{dUsv}),
and as a result the $\sd \times \si$ term vanishes. 
Note that a relatively small quantity $(\q + \qp)/4$ determines the change
in the computed $A_y$. Furthermore, we simplify the integral by 
approximating $\er$ of \eq{newc} by zero. The relative energy
is very low, $\er = E_{\rm d} + 2 E_n/3 - 3 (\p + \frac{\q}{2})^2/(4m)$
(with deuteron binding energy $E_{\rm d} = -2.22 \mev$), and
we have $\er < 0$ for the energy of interest ($E_n = 3 \mev$).
So the imaginary part vanishes. Since we do not include effective 
range corrections, we further neglect the energy dependence 
of the real part of \eq{newc}. It is necessary to reexamine this 
treatment within the framework of a Faddeev calculation that we 
shall advocate below. We thus find
\begin{widetext}
\be
\delta M^{\rm sv}_{\mjp,\mnp;\mj,\mn}(\qp,\q)
= \frac{i}{24 m^2} \, (3 C_0 + C_1 ) \int d\p \:
\bigl\langle \mjp , \mnp \bigl| \, \phiop(\p + \Delta/2) \,
\bigl[ (\sd - \si) \cdot \q \times \qp \bigr]
\, \phiop(\p-\Delta/2) \, \bigr| \mj , \mn \bigr\rangle \,.
\ee
\end{widetext}
Since $\sd$ commutes with the tensor operator $S_{12}(\phat)$, we
can move the operator $[ (\sd - \si) \cdot \q \times \qp ]$ to the 
right of $\phiop(\p-\Delta/2)$ and insert a one operator in deuteron 
spin space $\openone = \sum_{m_j''} | m_j'' \ra \la m_j'' |$. Using
\begin{multline}
\int d\p \: \la \mjp | \, \phiop(\p + \Delta/2) \, \phiop(\p-\Delta/2) 
\, | m_j'' \ra \\[1mm] 
= \delta_{m_j', m_j''} + \ord(\Delta^2) \,,
\end{multline}
we can neglect the momentum dependence of the charge form factor,
as well as the magnetic and quadrupole form factors of the deuteron.
With $\q \times \qp = q^2 \sin\theta \: \widehat{\bf y}$, we
have for the spin-violating boost corrections in operator form
\be
\delta M_{\rm sv} = i \, R \, (\sd - \si)^y = i \, 
\frac{q^2 \sin\theta}{24 m^2} \, (3 C_0 + C_1) \, (\sd - \si)^y \,.
\ee
Here we have for convenience combined all factors into the coefficient
$R$. Combining our results with Eq.~(\ref{dAygen}) leads to
\be
\delta A_y = i \, R \, 
\frac{\tr \bigl( \sd^y \, \si^y \, \mdag - M \, \sd^y \, \si^y + M - 
\mdag \bigr)}{\tr \bigl( M \mdag \bigr)} \,.
\label{dAyfinal}
\ee
The necessary spin matrix element follows from the Wigner-Eckert
theorem:
\be
\la \mjp | \, \sd^y \, | \mj \ra = i \, \bigl[
( 1 \, \mjp \, 1 \, 1 | 1 \, \mj ) + 
( 1 \, \mjp \, 1 \, -1 | 1 \, \mj ) \bigr] \,.
\ee

We are now in the position to study the impact on $A_y$. For the
$nd$ scattering amplitude $M$ in Eq.~(\ref{dAyfinal}) we use the
phase shifts from Kievsky {\it et al.}~\cite{Kievsky,KievskyPS}. 
These are based on the Argonne $v_{18}$ NN and the
Urbana 3N interaction for $J^P$ up to $7/2^+$ (from Table~2
in Ref.~\cite{Kievsky}) and on the Argonne $v_{14}$ NN interaction
for $7/2^-$ and $9/2 \leqslant J \leqslant 13/2$ (from Tables~I
and~II in Ref.~\cite{KievskyPS}). No parameters are adjusted.
As a check, we have reproduced the differential cross section
of Ref.~\cite{Kievsky}, which is in very good agreement with
experiment.

\begin{figure}[t]
\begin{center}
\includegraphics[scale=0.36,clip=]{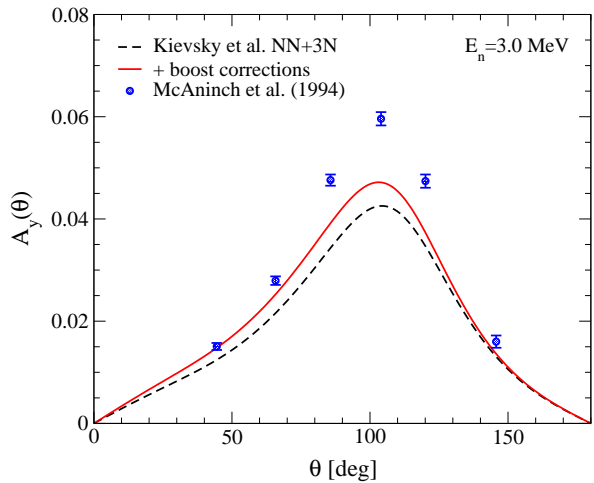}
\caption{(Color online) The $nd$ analyzing power $A_y$ for $E_n
= 3 \mev$ as a function of center-of-mass scattering angle $\theta$. 
The dashed curve is based on $nd$ phase shifts obtained from NN 
and 3N interactions~\cite{Kievsky,KievskyPS} (for details
see text). The solid curves include our results for the boost
corrections without distortion. The data is taken from
McAninch {\it et al.}~\cite{McAninch}.}
\label{Ay_ndE3}
\end{center}
\end{figure}

The effect of the spin-violating boost corrections on $A_y$ is shown 
in Fig.~\ref{Ay_ndE3} for $E_n = 3 \mev$ and in comparison to the data 
from McAninch {\it et al.}~\cite{McAninch}. We see that the influence 
of the spin-violating relativistic corrections is to increase
the computed value of $A_y(\theta)$ by about $10 \%$ at the peak.
This contribution is significant. It shows that relativistic effects 
may be relevant even at very low energies due to resonant
enhancements. However, this effect alone is too small to solve the 
$A_y$ puzzle. We therefore explore the effects of initial and
final state interactions with the nucleon that has been treated as a 
spectator so far, and finally, we discuss the energy dependence of
these boost effects.

We have considered the effects of boosting the interaction between
nucleons 13 and 23, while treating the nucleon 2 and 1 as a spectator.
The total momentum of the boosted pair is effectively $(\q +\qp)/4$.
When for instance the projectile neutron interacts with nucleon 2 
before interacting with nucleon 1, the total momentum of the 13 
nucleon-pair will be increased due to the attractive interaction 
between nucleons 2 and 3, and we expect our boost effect to be 
enhanced. We explore the size of a 23 interaction with a schematic
square-well $^1$S$_0$ potential of Ref.~\cite{BrownJackson}, which 
has a depth $V_0 = 13.4 \mev$ and range $R = 2.65 \fm$. Using
conventional NN interactions, we estimate the probability to find
two nucleons in a deuteron closer than $R$ to be about $50 \%$,
so that a 23 interaction can be followed by a 13 interaction about 
half of the time. If this occurs, the relative momentum inside
the well $\kappa$ is given by
\be
\kappa^2 = \frac{3 q^2}{4} + m V_0 = m \, \biggl( \frac{2 E_n}{3} + V_0 
\biggr) \,,
\ee
so that $\kappa \approx 0.6 \fmi$. These prescattering contributions
occur about half of the time, and thus the relevant average momentum 
is $\approx 0.3 \fmi$. This value is about three times larger than 
$| \q + \qp |/4 \approx 0.1 \fmi$. Therefore, we expect that distortion
effects will increase $A_y$ further. This simple estimate should be 
taken only as an assessment that initial state interactions can make
a large contribution to the boost effects.

With increasing energy, the resonant enhancement of the spin-violating
relativistic corrections decreases. This is both due to the effective
range $r_i$ (and the decrease of the S-wave phase shifts with increasing
energy), 
\be
\frac{1}{a_i} \to \frac{1}{a_i} - \frac{r_i \, m \, E_{\rm rel}}{2} \,,
\ee
and the impact of the imaginary part $i \sqrt{m \, \er}$. A detailed
study of the energy dependence of these boost effects is beyond the
scope of this paper and will be left to a future 
investigation~\cite{MS}.

\section{Summary and Future Steps}
\label{Concl}

In this paper, we have presented the first estimate of the effects
of relativistic boost corrections on the $nd$ analyzing power $A_y$.
We have focused on spin-violating relativistic corrections at order
$(Q/m)^2$, which can be important at low energies due to a resonant
enhancement from the large S-wave scattering lengths. Since boost 
corrections depend on the two-body cm momentum, the modern viewpoint
is to consider their effects as 3N interactions. We have used the
formalism of Ref.~\cite{rel1,rel2,rel3}, where it is straightforward 
to include relativistic corrections to order $(Q/m)^2$ without 
any new parameters. The relevant spin-violating
contribution to the $nd$ transition 
amplitude is given in \eq{dUsv}.

These corrections induce a $10 \%$ change in the computed value of 
the $nd$ analyzing power $A_y$ for laboratory energy $E_n = 3 \mev$.
This is a small, but significant contribution of the sign necessary 
to resolve the $A_y$ puzzle. Our result was estimated using the plane-wave 
impulse approximation, which leads to a transparent analytical 
calculation. The present study is clearly not complete. The 
effects of initial and final state interactions allow for additional
contributions. For instance, the effects of $\delta V_{12}$ would 
not vanish (as in the present calculation), if initial or final state  
interactions excite the deuteron. Faddeev calculations that include
distortions are therefore needed to conclude on the $A_y$ puzzle. The 
results presented here are mainly intended to stimulate the interest 
of the few-body community to include relativistic corrections in their 
complete solutions of the 3N problem.

For energies $E_n \gtrsim 30 \mev$, the predicted $A_y$ based on
microscopic NN and 3N interactions (without relativistic corrections)
is in very good agreement with experiment. Our present results are
not in contradiction to these findings, since the resonant 
enhancement of our spin-violating boost corrections decreases with 
energy. A detailed study of the energy dependence will be presented
in a future paper~\cite{MS}. In addition, future work will estimate 
the scaling to larger systems and the impact on the $A_y$ puzzle
in $n \, ^3$H scattering~\cite{MS}.

\acknowledgments

We thank Dick Furnstahl, 
Andreas Nogga and Rob Timmermans for useful discussions.
This work was supported in part by the US Department of Energy 
under Grant No.~DE--FG02--97ER41014 and the Natural Sciences and 
Engineering Research Council of Canada (NSERC). TRIUMF receives federal
funding via a contribution agreement through the National Research 
Council of Canada.

\end{document}